\documentclass[]{raa}           
\usepackage{graphicx,times}
\usepackage{natbib}

\providecommand{\bm}[1]{\mbox{\boldmath$#1$\unboldmath}}

\begin{document}

   \title{Discovery of six high-redshift quasars with the Lijiang 2.4m telescope and the Multiple
Mirror Telescope
% $^*$
%\footnotetext{\small $*$ Supported by the National Natural Science Foundation of China.}
}

 \volnopage{ {\bf 2012} Vol.\ {\bf 12} No. {\bf 9}, 1185--1190}
   \setcounter{page}{1}

   \author{Xue-Bing Wu\inst{1}, Wenwen Zuo\inst{1}, Qian Yang\inst{1}, Weimin Yi
      \inst{2},  Chenwei Yang\inst{3,4}, Wenjuan Liu\inst{3,4}, Peng Jiang\inst{3,4}, Xinwen Shu\inst{3,4}, Hongyan Zhou\inst{3,4}
   }

   \institute{ Department of Astronomy, Peking University, Beijing 100871, 
China; {\it wuxb@pku.edu.cn}\\
        \and
             Yunnan Astronomical Observatory, National Astronomical Observatories, Chinese Academy of Sciences,
             Kunmin 650011, China\\
	\and
%	  Center for Astrophysics, University of Science and Technology of China, Hefei 230026, China\\
Key Laboratory for Research in Galaxies and Cosmology, The University of Science
and Technology of China, Chinese Academy of Sciences, Hefei, Anhui, 230026, China\\
\and 
Polar Research Institute of China,
Jinqiao Rd. 451, Shanghai, 200136, China\\
\vs \no
   {\small Received 2012 June 12; accepted 2012 July 27 }
}

\abstract{Quasars with redshifts greater than 4 are rare, and can be used to probe the structure and evolution of the early universe. Here we report the discovery of six new quasars with $i$-band magnitudes brighter than 19.5 and redshifts between 2.4 and 4.6 from the YFOSC spectroscopy of the Lijiang 2.4m telescope in 
 February, 2012. These quasars are in the list of $z>3.6$ quasar candidates selected by using our proposed $J-K/i-Y$ criterion and the photometric redshift estimations from the SDSS optical and UKIDSS near-IR photometric data. Nine candidates were observed by YFOSC, and five among six new quasars were identified as $z>3.6$ quasars. One of the other three objects was identified as a star and  the other two were unidentified due to the lower signal-to-noise ratio of their spectra.  This is the first time that $z>4$ quasars have been discovered using a telescope in China. Thanks to the Chinese Telescope Access Program (TAP), the redshift of 4.6 for one of these quasars was confirmed by the  Multiple
Mirror Telescope (MMT)  Red Channel spectroscopy.  The continuum and emission line properties of these six quasars, as well as their central black hole masses and Eddington ratios, were obtained. 
}

   \authorrunning{X.-B. Wu et al. }            %author_head in even pages
   \titlerunning{Discovery of six high-redshift quasars}  % title_head in odd pages
   \maketitle

%% The author head (on even pages) and the title head (on odd pages) will be
%% automatically extracted from \author{} and \title{}. Whenever the title is too long,
%% you will be asked to supply a shorter one by inserting either \authorrunning{} or
%% \titlerunning{} before \maketitle. Anyway, you can specify your own heads.
%%
%%
%% Note: In the following text body of your manuscript, please note several differences from
%%       other major journals:
%% (1) \subsection{Please Capitalize the First Letter of Each Notional Word in Subsection Title}
%% (2) Please Capitalize the First Letter of Each Notional Word in all tables' captions

%
%________________________________________________ sections below
%
\section{Introduction}           %% first-level sections will be auto-capitalized
\label{sect:intro}

The number of known quasars has increased steadily in the past four decades since their discovery  in
 1963 (Schmidt 1963). In particular, a large number of quasars have 
been discovered in two large spectroscopic surveys, namely, the Two-degree 
Field (2dF) survey (Boyle et al. 2000) and the Sloan Digital Sky Survey (SDSS)
(York et al. 2000). 2dF mainly selected low redshift ($z<2.2$) quasar candidates with UV-excess 
(Smith et al. 2005) and has discovered more than 20,000 quasars (Croom et al. 
2004).  SDSS adopted
a multi-band optical color selection method for quasars mainly by
excluding the point sources in the stellar locus of the color-color 
diagrams (Richards et al. 2002) and has identified more
than 120,000 quasars (Schneider et al. 2010).  
90\% of SDSS quasars have low 
redshifts ($z<2.2$), though some 
dedicated methods were also proposed for finding high redshift quasars ($z>3.5$)
(Fan et al. 2001a,b; Richards et al. 2002). 

High-redshift quasars are rare, and those with redshifts 
greater than 4 represent only 1\%  in the total quasar population. In
the SDSS DR7 quasar catalog (Schneider et al. 2010), only 1248 (392) among  105783 quasars have redshifts
greater than 4 (4.5). Since these $z\sim4$ quasars exist when the universe is at age of 1.57 Gyr, they
can be used to probe the structure and evolution of the early universe (Smith et al. 1994; Constantin et al. 2002). In particular, the
absorption line spectra of these quasars can give valuable information on the nature
of intergalactic medium at high redshift.
However, discovering $z\sim4$ quasars is a big challenge  because they are 
fainter than the low redshift quasars due to their larger distances.  Moreover,
the Ly$\alpha$ emission lines for  $z\sim4$ quasars move to the red end of optical spectra,
making them hard to be distinguishable from stars due to similar optical
colors. Recently, Wu \& Jia (2010) 
proposed using the $Y-K/g-z$ criterion to select $z<4$ quasars and using the $J-K/i-Y$
criterion to select $z<5$ quasars with 
the SDSS optical and UKIDSS (UKIRT  Infrared Deep Sky Survey)\footnote{The UKIDSS project is defined in Lawrence et al. (2007). UKIDSS uses the UKIRT Wide Field Camera (WFCAM; Casali et al. 2007) and a photometric system described in Hewett et al. (2006). The pipeline processing and science archive are described in Hambly et al. (2008). } near-IR data based on a K-band excess technique (Warren et al. 2000; Hewett et al. 2006; Chiu et al. 2007; 
Maddox et al. 2008).
With these two criteria, we expect to obtain more complete quasar samples than previous ones.
Recent optical spectroscopic observations made by the GuoShouJing Telescope (LAMOST)
and  MMT have demonstrated the success of finding the missing quasars
with redshifts between 2.2 and 3 using the Y-K/g-z criterion (Wu et al. 2010a,b; Wu et al. 2011).
We also hope to discover some  $z\sim4$ quasars with the  J-K/i-Y criterion, which is expect to be
applicable for selecting the candidates of quasars with redshifts up to 5 (Wu \& Jia 2010).

In this letter, we report our discovery of six  new high redshift quasars from the spectroscopic observations with the Lijiang 2.4m telescope and MMT in February, 2012.   
The successful identifications of these high redshft quasars further demonstrate the effectiveness of using our newly proposed criteria for discovering the missing quasars including high-redshift ones.

% Authors can give a citation as `Michel et al. 1992'.
% You may also use \cite, \citep and \citet for citation, and use Table~1
% or Figure~1 and so forth. Using \ref and \label for cross-references of
% Tables/Figures is a good way in adjusting/adding/removing text, tables or
% figures.

\begin{figure}
   \centering
   \includegraphics[width=14.0cm, angle=0]{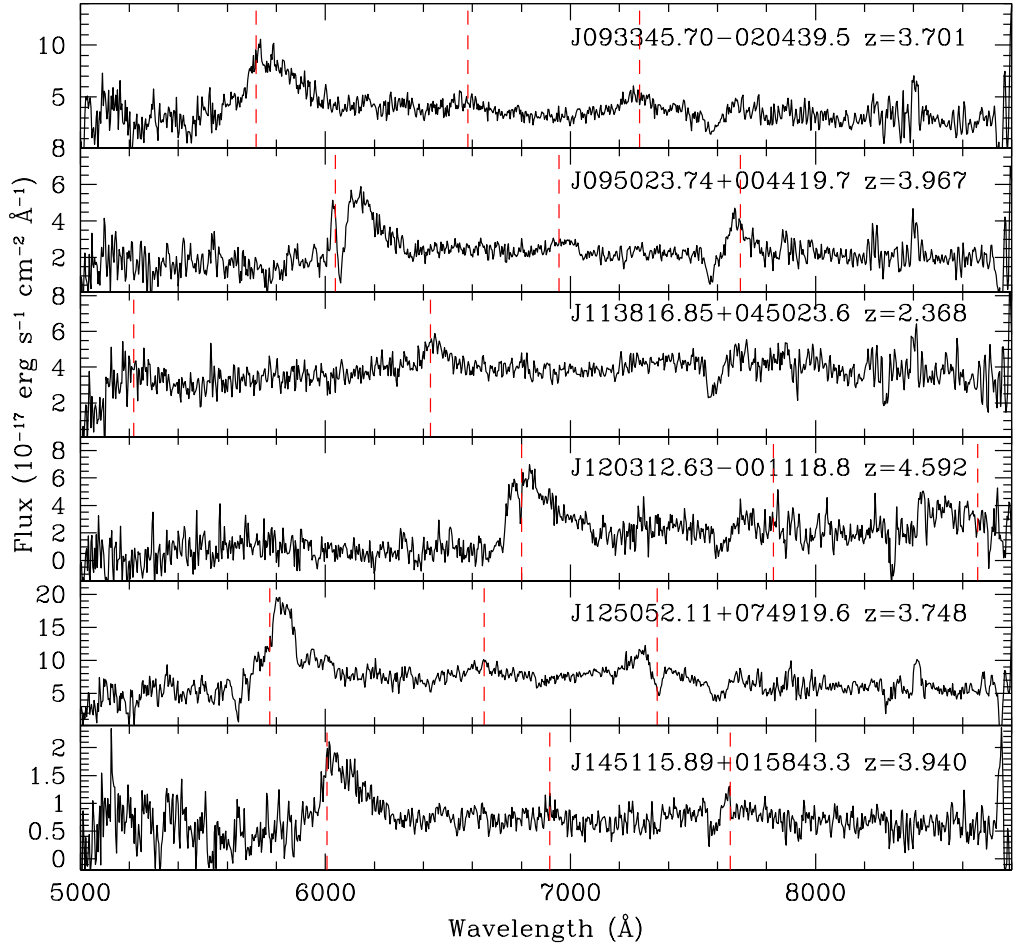}
  % \begin{minipage}[]{85mm}
   \caption{The YFOSC spectra of six new quasars. From the left to right, the red dashed lines mark the wavelengths of Ly$\alpha$, SiIV and CIV emission lines at the estimated redshift for five $z>3.6$ quasars, while for SDSS J113816.85+045023.6 they mark the wavelengths of CIV and CIII]. } 
%\end{minipage}
   \label{Fig1}
   \end{figure}

\section{Target Selection and Spectroscopic Observations}
\label{sect:Obs}

Richards et al. (2009) presented a catalog of about 1million quasar candidates selected from the SDSS DR6 photometric data  using Bayesian
methods. Photometric redshifts for these candidates were also provided
based on the SDSS $urgiz$ magnitudes. From this catalog we selected all unidentified candidates with the photometric redshift greater than 3.6, the photometric redshift probability larger than 0.6  and the $i$-band magnitude brighter than 19.5. Then we
cross-matched them with the UKIDSS Large Area Survey (LAS) DR7 catalog using a positional offset of 3 arcsec to find the closest counterparts.  From this sample with both SDSS $ugriz$ data and UKIDSS $YJHK$ data, we adopted our
 $ J-K/i-Y$ criterion (Wu \& Jia 2010), namely, $J-K>0.45(i-Y) +0.48$ (where $YJK$ are the Vega magnitudes and $i$ is the AB magnitude) , to make further selection of  $z\sim4$ quasar candidates. We also used our
own program to estimate the photometric redshifts of these candidates with SDSS and UKIDSS 9-band
photometric data (Wu \& Jia 2010; Wu, Zhang \& Zhou 2004), and excluded the sources whose photometric redshifts estimated from the 5-band SDSS photometric data
in Richards et al. (2009) are inconsistent with ours. After these procedures we obtained a final list
of about 20 high-redshift ($z>3.6$) quasar candidates. 

The spectroscopic observations were carried out on February 26-28, 2012, with the Yunnan Faint Object Spectrograph and Camera (YFOSC) instrument of
the Lijiang 2.4m telescope in Yunnan Astronomical Observatory. Due to the cloudy weather, nine candidates
were observed with YFOSC using  a low resolution grism with the central wavelength around $6500\AA$ , the spectral resolving power of 870, and a long slit with of 2.5$''$ width. The typical seeing is around 2$''$. In Table~1 we summarize the details of the observations for these 9 candidates. Six of them were idenfied as
quasars, one as a G-type star and two as unidentified due to the lower signal-to-noise ratios of their spectra. 

\begin{table}
\bc
\begin{minipage}[]{100mm}
\caption[]{Parameters of 9 objects observed by YFOSC}\end{minipage}
\setlength{\tabcolsep}{1pt}
\small
 \begin{tabular}{ccccccccccccc}
  \hline\noalign{\smallskip}
Name& Date& Exposure&$u$& $g$& $r$& $i$& $z$& $Y$& $J$& $H$& $K$& Result\\
(SDSS J)& &($s$)&&&&&&&&&&\\
  \hline\noalign{\smallskip}
075733.86+190403.1&2012-02-26&2700&21.37&20.40&19.45&19.00&18.53&17.88&17.02&16.63&15.68&low S/N\\
085203.84+020437.7&2012-02-27&6000&21.67&20.99&19.69&19.09&18.66&17.82&17.47&16.66&15.72&low S/N\\
092740.04-023347.5&2012-02-28&4200&21.01& 20.50&19.55&19.08&18.86&18.42&18.01&17.11&16.42&G star\\
093345.70-020439.5&2012-02-27&3600&23.77&20.72&19.55&19.47&19.41&19.21&18.66&18.46&17.88&quasar \\
095023.74+004419.7&2012-02-27&6600&23.97&20.97&19.75&19.48&19.36&18.83&18.40&17.76&17.25&quasar\\
113816.85+045023.6&2012-02-27&6000&21.26&20.70&19.67&19.27&19.06&18.30&17.96&16.90&16.46&quasar\\
120312.63-001118.8&2012-02-28&5400&25.38&22.34&20.29&19.14&18.95&18.32&18.00&17.19&16.64&quasar\\
125052.11+074919.6&2012-02-26&5400&23.56&20.08&18.75&18.63&18.43&18.14&17.35&16.81&16.15&quasar \\
145115.89+015843.3&2012-02-27&4800&23.72&20.42&19.30&19.23&19.09&18.61&18.02&17.56&16.92&quasar \\
  \noalign{\smallskip}\hline
\end{tabular}
\ec
%% place \tablecomments and \tablerefs below \end{center| and \end{center}:
%% you may leave the table-width parameter to editors or set to your actual size
\tablecomments{0.86\textwidth}{The $ugriz$ magnitudes are given in AB system 
and the $YJHK$ magnitudes are given in Vega system.}
\end{table}

The spectra of six new quasars, after the flat-field correction and both wavelength and flux calibrations, are shown in Fig.1. The strongest emission lines for five $z>3.6$ quasar are Ly$\alpha$ lines, while for SDSS J113816.85+045023.6 the strongest line around $6400\AA$ is CIII]. The redshifts for these quasars are the average values given mostly by the Ly$\alpha$ 
and CIV lines for five $z>3.6$ quasar and by the  III] and CIV lines for SDSS J113816.85+045023.6.

Thanks to the Chinese Telescope Access Program\footnote{http://info.bao.ac.cn/tap/},  SDSS J120312.63-001118.8 was also observed  with the Red Channel Spectrograph on 
the MMT 6.5m telescope\footnote{Observation reported here was obtained at the MMT Observatory, a joint facility of the Univeristy of Arizona and the Smithsonian Institution.} at Mt. Hopkins, Arizona, USA on Feb. 29, 2012, with a wavelength coverage of 5100-8600$\AA$ 
and a spectral resolution of 1.6\AA. It was observed twice, with the exposures of 10 minutes and 15
minutes respectively. The spectrum was processed using the IRAF Echellette package and is shown in Fig. 2.
The average redshift estimated from Ly$\alpha$ and SiIV (1400\AA) emission lines is 4.601$\pm$0.008, 
consistent with the result (z=4.592$\pm$0.048) obtained from the YFOSC observation.

\begin{figure}
   \centering
  \includegraphics[width=14.5cm, angle=0]{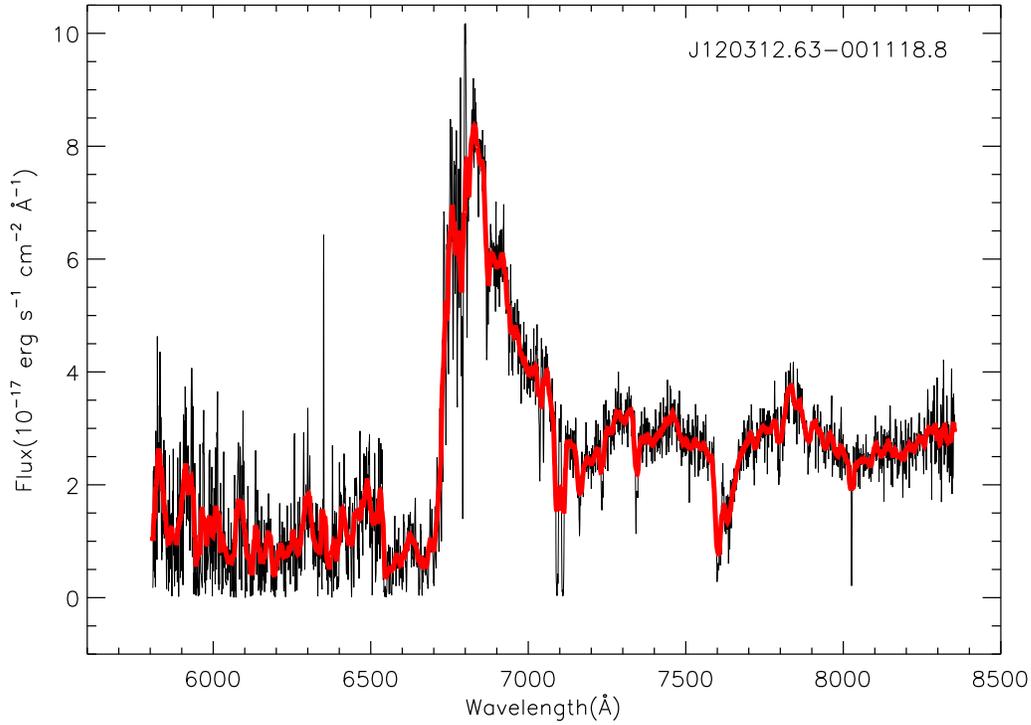}
  % \begin{minipage}[]{85mm}
   \caption{The MMT Red Channel spectrum of SDSS J120312.63-001118.8. The average redshift estimated from Ly$\alpha$ and SiIV (1400\AA) emission lines is 4.601. The
red line refer to the smoothed spectrum with a binsize of 8\AA.} 
%\end{minipage}
   \label{Fig2}
   \end{figure}

\section{Spectral analyses and properties of six high-redshift quasars}

The redshift corrected rest-frame YFOSC spectra of six quasars are first corrected for the Galactic extinction using the extinction map of 
Schlegel et al. (1998). They are then fitted with an IDL code MPFIT \citep{Markwardt09}.
We fit the spectra with the  pseudo-continuum model consisting of the featureless nonstellar continuum and FeII emissions. The continuum is assumed to be a power-law, so two
free parameters (the amplitude and the slope) are required. 
The UV FeII template (Vestergaard \& Wilkes 2001; Tsuzuki et al.  2006) is convolved with a velocity dispersion and shifted with a velocity, assuming the line width of 
FeII lines are same with emission lines in the corresponding wavelength range.
During the fitting, the amplitude and slope of the power-law continuum and the amplitude, velocity shift and
broadening width of the FeII emission, are set to be free parameters. The initial value of the power-law
continuum is obtained by fitting a simple power law to the data in the chosen windows, which are free from 
emission-line contamination. The initial value of broadening width of the FeII emissions is set to be the 
line width of the strong emission line CIV. Then we use the pseudo-continuum model to fit a set of continuum windows 
with strong FeII emissions but no other emission lines, 
as mentioned in Hu et al. (2008), slightly adjusted interactively for each individual spectrum
in order to avoid broad absorption features or extended wings of emission lines. 

After constructing the pseudo-continuum, the CIV line should be fitted with two Gaussians, one for the narrow component and another for the broad component. However, except SDSS J095023.74+004419.7,  the spectra of other five quasars have low signal-to-noise ratio, so we used only one Gaussian to fit the CIV emission line. Absorption features are evident in the spectra of four quasars, and one more negative Gaussian was added in the fitting.
We measure the Full Width at Half Magnitude 
of CIV line (FWHM$_{\rm CIV}$), luminosity at 1350 \AA ($L_{1350}$) from the spectra (except for 
SDSS J113816.65+004419.7, where 1350$\AA$ is not within the wavelength coverage, we estimate the luminosity at 1500$\AA$  
instead). The black hole mass is estimated based on FWHM$_{\rm CIV}$ and $L_{1350}$  with Eq. (7) in Vestergaard \& Peterson (2006)(see also Kong et al. 2006). Using a scaling between $L_{1350}$ and bolometric luminosity 
$L_{\rm bol}$, $L_{\rm bol}=4.62L_{1350}$, we estimated the bolometric luminosity for the six quasars. 
Based on the estimated black hole mass and bolometric luminosity, we also estimated their Eddington ratios 
($R_{\rm EDD}$). 
The results are summarized in Table~\ref{mbh}. 
Although the uncertainties of these values are probably quite large due to the low spectral quality and
the unusual properties of CIV, we noticed that the overall properties of these six quasars are consistent with those of typical SDSS quasars at high redshift (Shen et al. 2011).

\begin{table}
\bc
\begin{minipage}[]{100mm}
\caption[]{Spectral parameters and black hole masses of six new quasars\label{mbh}}\end{minipage}
\setlength{\tabcolsep}{2.5pt}
\small
 \begin{tabular}{ccccccccccccc}
  \hline\noalign{\smallskip}
Name &  Redshift & FWHM(CIV) & $\log (L_{1350})$ & $\log (M_{BH})$ &
$\log (L_{\rm bol})$ & $\log R_{\rm EDD}$ \\
 (SDSS J)&  & ($km~s^{-1}$) & ($erg~s^{-1}$) &($M_\odot$) &($erg~s^{-1}$) &  \\
  \hline\noalign{\smallskip}
093345.70-020439.5& 3.701$\pm$0.011 &  6749 &  46.457&  9.621&   47.122  &
  -0.588  \\
095023.74+004419.7& 3.967$\pm$0.035 &  4500 &  46.300&  9.185&   46.964  &
  -0.310  \\
113816.85+045023.6& 2.368$\pm$0.011 &  5144 &  46.000&  9.118&   46.664  &
  -0.543  \\
120312.63-001118.8& 4.592$\pm$0.048 &  4865 &  46.431&  9.323&    47.096  &
  -0.316  \\
125052.11+074919.6& 3.748$\pm$0.030 &  5424 &  46.805&  9.615&   47.469  &
  -0.235  \\
145115.89+015843.3& 3.940$\pm$0.006 &  4507 &  45.682&  8.859&   46.346  &
  -0.602  \\
  \noalign{\smallskip}\hline
\end{tabular}
\ec
%% place \tablecomments and \tablerefs below \end{center| and \end{center}:
%% you may leave the table-width parameter to editors or set to your actual size
%\tablecomments{0.86\textwidth}{
%Black hole masses estimated with Eq. (7) in \cite{Vestergaard06}.}
\end{table}

\section{Discussion}
\label{sect:discussion}
A complete quasar sample is crucial for studying the large scale structure of the universe. The current available quasar samples
are mostly  biased towards low redshifts ($z<2.2$)  and more efforts are needed to find quasars at high redshift.
 Wu \& Jia (2010) proposed to obtain a large complete quasar sample with redshifts up to five by combining the  $J-K/i-Y$ criterion with the $Y-K/g-z$ criterion to select quasar candidates. Some recent
optical spectroscopic observations  have demonstrated the success of finding the missing quasars
with redshifts between 2.2 and 3 using the $Y-K/g-z$ criterion (Wu et al. 2010a,b; Wu et al. 2011). Our discovery of six  high redshift quasars (five with $z>3.6$) from the spectroscopic observations with the Lijiang 2.4m telescope  and MMT further demonstrates the effectiveness of using the  $ J-K/i-Y$ criterion for discovering  quasars with redshifts up to five. Moreover, the identification of five quasars with $z>3.6$ from nine candidates with photometric
redshift larger than 3.6 also confirms the robustness of the photometric redshifts estimated by the SDSS and UKIDSS photometric data.
We noticed that two among our five $z>3.6$ new quasars do not meet the SDSS $gri$ or $riz$ selection critrion for
$z>3.6$ quasars (Fan et al. 2001a,b; Richards et al. 2002), which suggests that about 40\% of such quasars may be missed in the SDSS spectroscopic survey. This obviously needs to be confirmed by future observations of a large sample of $z>3.6$ quasars.

Our identification of a $z=4.6$ quasar demonstrates that $z>4$ quasars can be identified with the 2-meter size telescopes in China. We hope more high redshift
quasars will be discovered by the future LAMOST quasar survey (Wu 2011), which is aiming at discovering 0.3-0.4 million quasars from 1 million quasar candidates
with $i<20.5$, by taking the advantages of 4000 fibers and 5 degree field of view of LAMOST (Su et al. 1998; Zhao et al. 2012).    
The new quasar selection criteria, such as those based on SDSS, UKIDSS and the Wide-field Infrared Survey Explorer (WISE; Wright et al. 2010) data (Wu et al. 2012), will be applied for selecting quasar candidates in the
LAMOST quasar survey. This will hopefully provide the largest quasar sample in the next five years for further studies of AGN physics, large scale structure and cosmology.

\normalem
\begin{acknowledgements}
We thank the referee for a constructive report and Jianguo Wang, Cheng Hu and Zhaoyu Chen for great helps on the spectral analysis.
This work was supported by the National 
Natural Science Foundation of China  (11033001). 
We acknowledge the use of Lijiang 2.4m telescope and  the MMT 6.5m telescope, as well as
the archive data from SDSS and UKIDSS. 
This research uses data obtained through the Telescope Access Program (TAP), which is funded by the National Astronomical Observatories, Chinese Academy of Sciences, and the Special Fund for Astronomy from the Ministry of Finance. Funding for the SDSS and SDSS-II has been provided by the Alfred P. Sloan Foundation, the Participating Institutions, the National Science Foundation, the U.S. Department of Energy, the National Aeronautics and Space Administration, the Japanese Monbukagakusho, the Max Planck Society, and the Higher Education Funding Council for England. The SDSS Web Site is http://www.sdss.org/.

\end{acknowledgements}

%\appendix                  %%appendicial material is supported
%\section{This shows the use of appendix}
%A postscript file is actually an ASCII text file (you may even edit it).
%However, you need to transfer a PDF file or any compressed or packaged
%file in binary mode when using FTP.

%\label{lastpage}

%\newpage
%-------------------------------------------------------------

\end{document}